\documentclass[preprint,12pt]{elsarticle}

\usepackage{amssymb}
\usepackage{subfigure}

\journal{Nuclear Physics A}

\begin{document}

\begin{frontmatter}



\title{Study of $^{9}$Be+$^{12}$C elastic scattering at energies near the Coulomb barrier}


\author[usp]{R. A. N. Oliveira\corref{cor1}\fnref{label1}}
\ead{rnegrao@dfn.if.usp.br} \fntext[label1]{Tel.: +55 11 3091 7072;
fax: +55 11 3031 2742.} \cortext[cor1]{Corresponding Author}
\address[usp]{Instituto de F\'isica, Universidade de S\~ao Paulo, CEP
66318, 05315-970, S\~ao Paulo, SP, Brazil.}
\author[usp]{N. Carlin}

\author[usp]{R. Liguori Neto}

\author[usp]{M. M. de Moura}

\author[usp]{M. G. Munhoz}

\author[michigan]{M. G. del Santo}
\address[michigan]{National Superconducting Cyclotron Laboratory, Michigan State
University, East Lansing, Michigan.}
\author[usp]{F. A. Souza}

\author[usp]{E. M. Szanto}

\author[usp]{A. Szanto de Toledo}

\author[usp]{A. A. P. Suaide}


\begin{abstract}

In this work, angular distribution measurements for the elastic
channel were performed for the $^{9}$Be+$^{12}$C reaction at the
energies E$_{Lab}$=13.0, 14.5, 17.3, 19.0 and 21.0 MeV, near the
Coulomb barrier. The data have been analyzed in the framework of the
double folding S\~ao Paulo potential. The experimental elastic
scattering angular distributions were well described by the optical
potential at forward angles for all measured energies. However, for
the three highest energies, an enhancement was observed for
intermediate and backward angles. This can be explained by the
elastic transfer mechanism.

\end{abstract}

\begin{keyword}
$^{9}$Be+$^{12}$C \sep Elastic Scattering \sep S\~ao Paulo Potential


\PACS 24.10

\end{keyword}

\end{frontmatter}




\section{Introduction}
\label{introduction}

In the last few years, nuclear reactions involving weakly bound
nuclei became a subject of interest due to the observation of flux
enhancement for processes like nucleon transfer and breakup. Through
the study of these processes \cite{1,3,I6}, it is possible to obtain
information about nuclear structure, such as single-particle states
and nuclear cluster structure, as well as information about the
influence of continuum states in the nuclear reaction dynamics
\cite{4,5,6}. Additionally, the investigation on how these
properties change from the stability line to regions far from the
stability valley can also be addressed.

In this context, the elastic scattering measurement and coupled
channel analysis \cite{I9} are very important tools to investigate
nucleon transfer and breakup, as they appear as competing mechanisms
in the reproduction of the measured angular distributions.

In this work, elastic scattering cross section measurements were
performed for the ${^9}$Be+$^{12}$C reaction to study anomalies
\cite{I9,7,8} in the extracted optical parameters values and the
contribution of inelastic channels, transfer and compound nucleus
formation in the elastic scattering process. The experimental data
show an enhancement in the elastic cross sections at intermediate
and backward angles. This behavior is typically observed in systems
where projectile and target present the same core structure
\cite{7,8,I7}. This effect can be understood in terms of a ${^3}$He
transfer process, assuming that $^{12}$C has a ${^3}$He+$^{9}$Be
cluster structure.

The elastic scattering angular distributions were analyzed in a four
steps procedure. Distinctively of previous works, our experimental
elastic scattering data have been compared to optical model
predictions using a empirical double folding potential \cite{9}. In
this process, the normalization parameters for the real and
imaginary potentials were adjusted to describe the data at forward
angles ($\theta_{CM}$$\leq$80$^{\circ}$). The normalization of the
real part of the potential shows a decrease as a function of the
bombarding energy, and the normalization of the imaginary part of
the potential is approximately constant. During the following two
steps, we investigate the importance of the coupling to inelastic
and transfer channels respectively, and finally in the fourth step
we analyse the compound elastic contribution.


\section{Experimental Setup}
\label{setup}

The experiment was performed at the University of S\~ao Paulo
Physics Institute. The ${^9}$Be beam was delivered by the 8UD
Pelletron accelerator with energies E$_{Lab}$=13.0, 14.5, 17.3, 19.0
and 21.0 MeV (E$_{CM}$=7.4, 8.3, 9.9, 10.8 and 12.0 MeV
respectively) and hit a 40 $\mu$g/cm$^2$  thick $^{12}$C target. The
charged particles produced in the ${^9}$Be+$^{12}$C reaction were
detected by means of 13 triple telescopes \cite{10} separated by
$\Delta\theta$ = 10$^{\circ}$ in the reaction plane, which covered
the angular range from $\theta$ = 10$^{\circ}$ to $\theta$ =
140$^{\circ}$.

\begin{figure}[!h]

   \centerline{
   \subfigure[]{\includegraphics[scale=0.3]{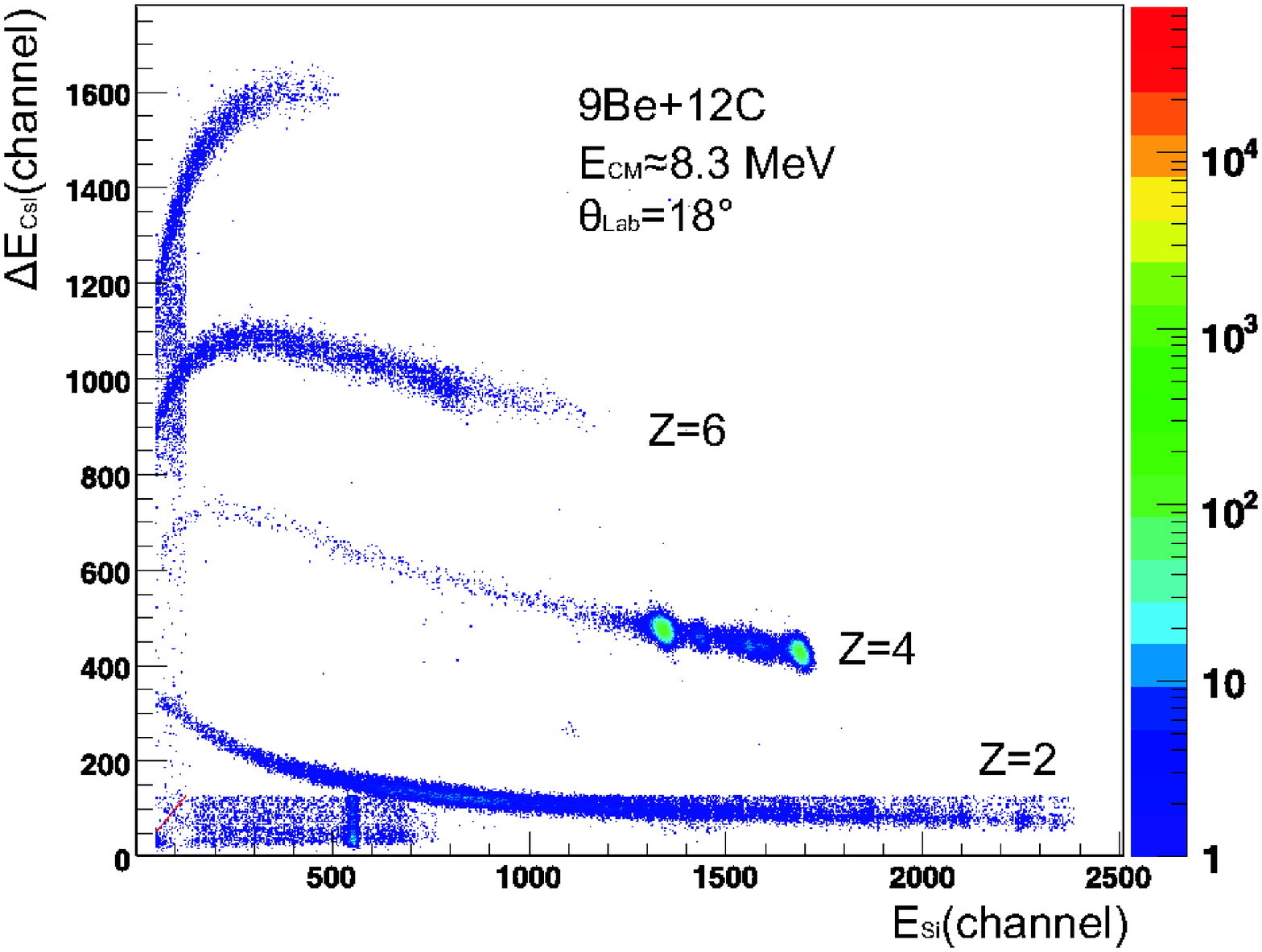}}
   \hfil
   \subfigure[]{\includegraphics[scale=0.3]{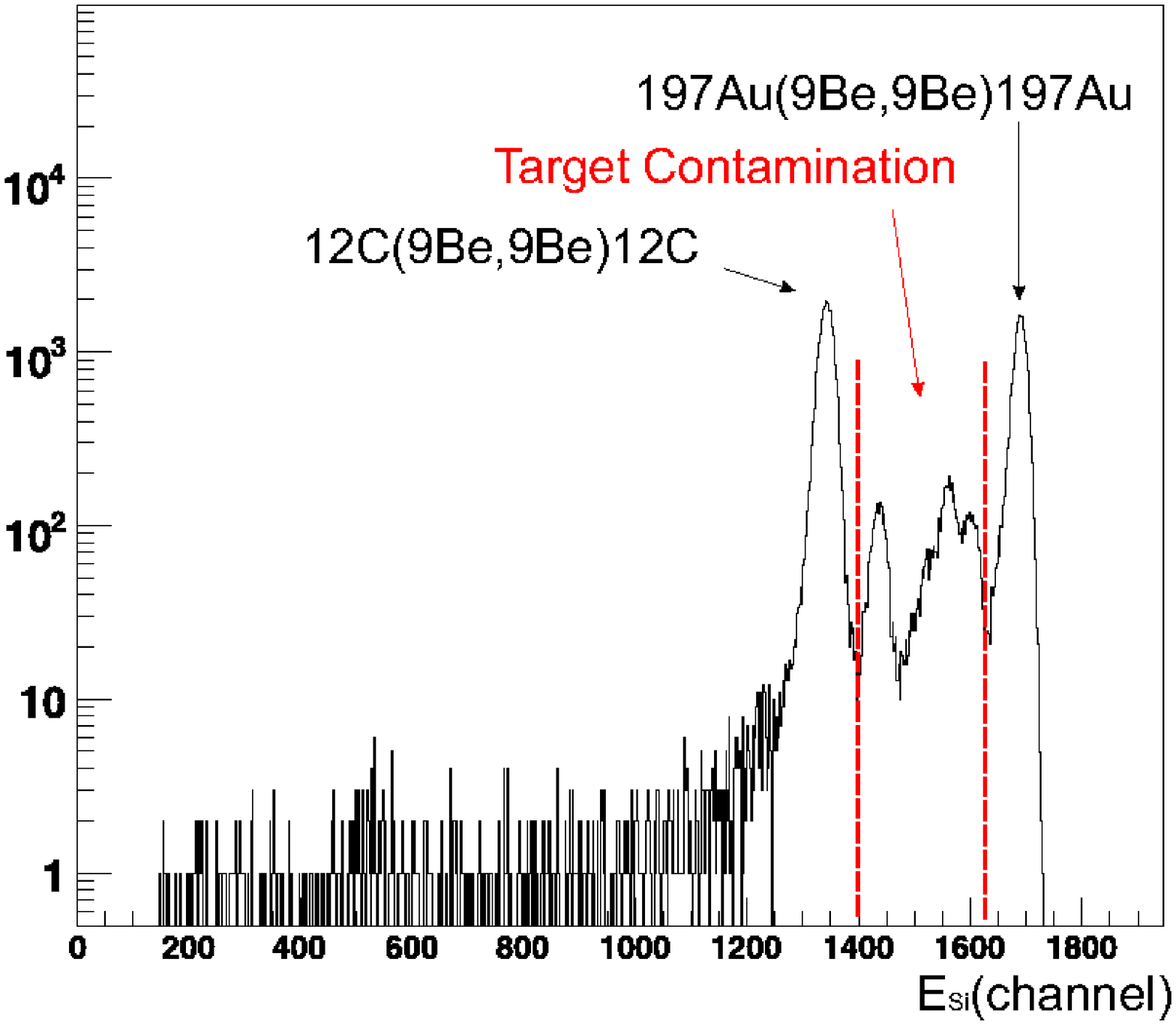}}
   }

\caption{(a) Two-dimensional spectra of $\Delta$E$_{gas}$ (energy
measured by the Gas detector) vs E$_{Si}$ at E$_{CM}$=8.3 MeV and
(b) Z=4 events projection on the energy axis (Color Online).}
\label{biparametrico}
\end{figure}

The triple telescopes were composed of an ionization chamber, filled
with 20 {\it torr} of isobutane gas, followed by a 150 $\mu$m
silicon detector and a 40 mm CsI scintillator crystal. The
identification of the elastic events was done by means of
two-dimensional spectra like the one shown in Fig.
\ref{biparametrico}-a. The elastic yields were obtained by
projecting the Z = 4 region on the E$_{Si}$ axis (energy measured by
the Silicon detector) and identifying the elastic processes as shown
in Fig. \ref{biparametrico}-b.

As ilustraded in this figure, we can see a considerable number of
events due to target contamination. In order to subtract the
contributions in the energy spectra from this target contamination,
we have assumed Rutherford scattering for the involved cross
sections. This is a good approximation in our experimental
conditions.

The uncertainties in the differential cross section were estimated
considering the statistical uncertainty in the yield and the
systematic uncertainty of 5\% in the target thickness.


\section{Data Analysis and Discussion}
\label{Analysis}

\subsection{Optical Model Calculations}

In the present work, the experimental data were analyzed using the
FRESCO code \cite{I4} in the framework of the S\~ao Paulo potential
(SPP) \cite{9} due to the recent success of this approach in
describing nuclear reactions involving weakly bound nuclei
\cite{I10,I11}. Clearly, there are other options for the scattering
potential, as shown in recent works by A. T. Rudchick {\it et. al.}
\cite{8} and J. Carter {\it et. al.} \cite{7} who have used a
Woods-Saxon shape and a double folding potential, respectively, and
the effects of coupled channels to describe the ${^9}$Be+$^{12}$C
scattering data.

For the S\~ao Paulo potential, the radial dependence is a folding
potential and the energy dependence takes non-local effects into
account, in the form
    \begin{equation}
    V(R,E)=V_F(R)\exp\left(-4\beta^2\right)\hspace{2mm},
    \end{equation}
where $\beta$=v/c, $v$ is the local relative velocity between the
two nuclei and $V_F(R)$ is a folding potential obtained by using the
matter distributions of the nuclei involved.

In detail, the folding potential depends on the matter densities in
the form

\begin{equation}
V_F(R)=\int
\rho(\overrightarrow{r}_1)\rho(\overrightarrow{r}_2)v_{nn}(\overrightarrow{R}-\overrightarrow{r}_1-\overrightarrow{r}_2)d^3r_1d^3{r_2}
\end{equation}
where
$v_{nn}(\overrightarrow{R}-\overrightarrow{r}_1-\overrightarrow{r}_2)$
is a physical nucleon-nucleon interaction given by

\begin{equation}
v_{nn}(\overrightarrow{R}-\overrightarrow{r}_1-\overrightarrow{r}_2)=V_0\delta(\overrightarrow{R}-\overrightarrow{r}_1-\overrightarrow{r}_2)
\end{equation}
with $V_0=-456$ MeVfm${^3}$ and the usage of a delta function
corresponds to the zero range approach. Extensive systematics were
performed in Ref. \cite{9}, to provide a good description of matter
and charge distribution.

\begin{figure}[htb!]
\centering
\includegraphics[scale=0.65]{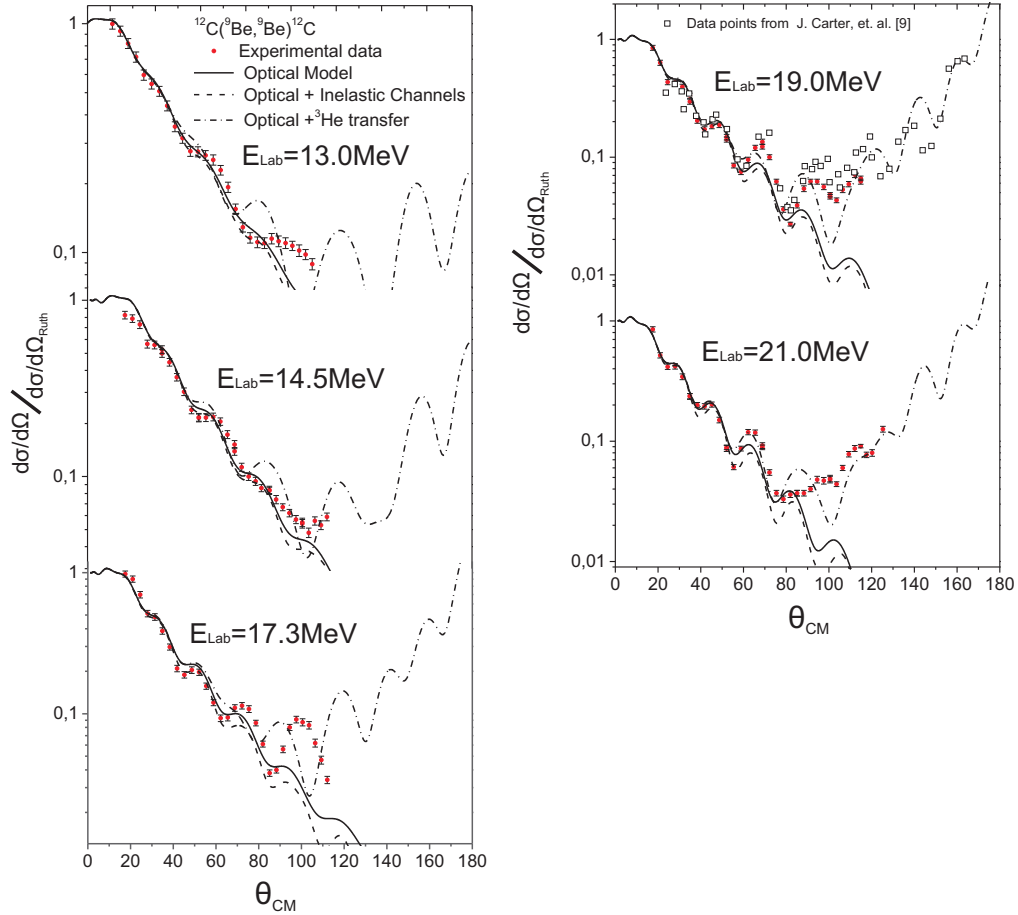}
\caption{Angular distributions for the $^9$Be + $^{12}$C system at
E$_{Lab}$ = 13.0, 14.5, 17.3, 19.0 and 21.0 MeV. The solid lines
correspond to optical model fits, the dashed and dash dotted lines
represent the same optical potential including the inelastic and
elastic transfer mechanisms respectively (Color Online).}
\label{elastico}
\end{figure}

In the present work, we adopt the matter and charge diffuseness
$a_m= 0.53$ fm and $a_c=0.56$ fm respectively, and the matter and
charge distribution radius given by $R_M = 1.31A^{1/3}-0.84$ and
$R_C = 1.76Z^{1/3}-0.96$ respectively. In this first step of the
analysis, we considered the real and imaginary potential
normalizations as free parameters that are adjusted in order to
describe the forward angles of the angular distribution
($\theta_{CM}$ $\leq$ 80$^{\circ}$). In this case the nuclear
potential is given by the equation
    \begin{equation}
    V_{SPP}=(N_r+iN_i)V_F(R)\exp\left(-4\beta^2\right)\hspace{2mm},
    \label{sppotential}
    \end{equation}
where the normalization coefficients $N_r$ and $N_i$ are reaction
energy dependent. The results for the first step are shown as solid
lines in Fig. \ref{elastico}. We observe a good agreement for the
forward angular region. However, a pronounced disagreement at
backward angles for 17.3, 19.0 and 21.0 MeV, is an indication of the
importance of other reaction mechanisms, not taken into account in
the optical potential. For 13.0 and 14.5 MeV, the optical model
description is reasonable. Finally, Fig. \ref{elastico} also shows
an angular distribution for E$_{Lab}$=19.0 MeV extracted from Ref.
\cite{7}. These data are in good agreement with our data and with
the optical model calculation at forward angles.

\begin{figure}[h!]
\centering
\includegraphics[scale=0.8]{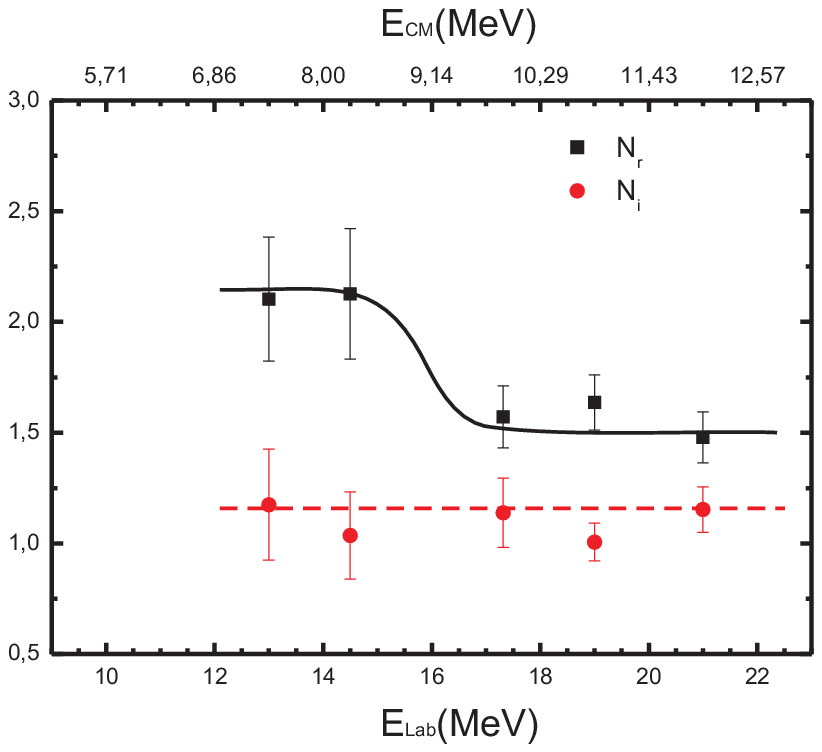}
\caption{Best values for N$_r$ and N$_i$ as a function of the
bombarding energy obtained from fits using the S\~ao Paulo potential
(Equation \ref{sppotential}) for the $^{9}$Be + $^{12}$C system. The
lines are just to guide the eyes (Color Online).} \label{nrni}
\end{figure}

The $N_r$ and $N_i$ energy dependence is depicted in Fig.
\ref{nrni}. The uncertainties are determined by a $\chi^2$ analysis.
One can notice that N$_r$ increases in the vicinity of the Coulomb
barrier energy (E$_{CM}$ $\approx$ 9 MeV \cite{8,I5}). The values of
N$_i$ are approximately constant for all energies, and they are in
agreement with the results obtained in Ref. \cite{I1}.

The increase of the N$_r$ parameter near the Coulomb barrier
suggests the presence of a threshold anomaly \cite{I2,I9}. However,
no strong statement about this anomaly can be made due to the
constant behavior of N$_i$.

The results from the first step of the analysis were used to perform
the coupling of the inelastic and transfer channels presented in the
next sections.

\subsection{Inelastic Channels}

With the optical potentials previously obtained, we are able to take
into account the effects of other channels. The second step of the
analysis consisted in calculating the effects of inelastic channels
in the theoretical elastic cross section, by considering 5/2${^-}$
and 7/2${^-}$ states of $^9$Be and 2${^+}$ of $^{12}$C.

In our study the transitions to excited states of $^9$Be and
$^{12}$C were calculated using the rotational model approach, where
the coupling interaction V$_{\lambda}$({\bf r}) of the multipole
$\lambda$ is

\begin{equation}
 V_{\lambda}({\bf r})=-\delta_{\lambda}\frac{dV({\bf r})}{dr}
\end{equation}

The Coulomb excitations are included by considering deformations on
the charge distributions in the form

\begin{equation}
V_{\lambda}^{C} = M(E\lambda)\frac{\sqrt{4\pi}e^2}{2\lambda+1}
\left\{
\begin{array}{rcl}
r^{\lambda}/r_c^{2\lambda+1}&r\leq r_c\\
1/r^{\lambda}&r>r_c\\
\end{array}
\right. \label{IN8}
\end{equation}

where

\begin{equation}
M(E\lambda)=\sqrt{(2J_{i}+1)B(E\lambda;J_{i}\to
J_{f})}\hspace{0.2cm}. \label{IN11}
\end{equation}

Therefore, to take into account the effects of inelastic process in
the theoretical elastic cross section, we take from the literature
\cite{7,8,11} the nuclear deformation parameters $\delta_{\lambda}$
and the reduced transition probabilities
B(E$\lambda$:J$_{i}$$\to$J$_{f}$), in order to calculate the
deformations on the symmetrical central potential. The parameters
for each channel included in the calculations and the reduced
transition probabilities are presented in table \ref{tabela1}.

\begin{table}[h]
\begin{centering}
\caption{Nuclear deformation parameters used in the inelastic
calculations.}
\begin{tabular}{p{2cm} p{2cm} p{2 cm} p{2cm} p{2cm} }
  \hline\hline
  Nucleus & Transition & $\lambda$ & B(E$\lambda$) & $\delta_{\lambda}$ \\
  \hline
   ${^9}$Be &  $\frac{3}{2}^-$ $\to$ $\frac{5}{2}^-$ &  2  & 46.0$\pm$0.5  &  2.4 \\
\\
   ${^9}$Be &  $\frac{3}{2}^-$ $\to$ $\frac{7}{2}^-$ &  2  & 33$\pm$ 1 &  2.4   \\
\\
   $^{12}$C &  $0^+$ $\to$ $2^+$ &  2  & 42$\pm$ 1  &  1.52  \\
\\
\hline\hline
\end{tabular}\\
\label{tabela1}
\end{centering}

\end{table}

The results are shown as dashed lines in Fig. \ref{elastico}. In
general, we observed that the inclusion of these channels decreases
the theoretical cross section when compared to the results obtained
only with the optical potential.

Comparing with the experimental data, we observed that the curves
show a good description at forward angles. For intermediate and
backward angles, the theoretical prediction underestimates the cross
section. This demonstrates that the inclusion of these inelastic
channels is not sufficient to explain the experimental results at
intermediate and backward angles.

\subsection{$^{3}$He Cluster Transfer}

In order to improve the description of experimental elastic
distributions at intermediate and backward angular region, a
coupling to the $^{3}$He transfer was included in the third step of
the analysis.

\begin{table}[h]
\begin{centering}
\caption{Optical parameters for DWBA calculations and spectroscopic
factor for a A=C+v system.} \label{tabela2}
\begin{tabular}{p{3cm} p{1.75cm} p{1.5cm} p{1.5cm} }
  \hline\hline
  Systems & V$_{r}$(MeV) & r$_{r}$(fm) & a$_{r}$(fm) \\
  \hline
\\
  ${^9}$Be+$^{9}$Be$^{a}$ &  189.3  &  1.0  & 0.63  \\
\\
  ${^3}$He+${^9}$Be &  55.8$^{b}$  & 1.35 & 0.65    \\
\\ \hline\hline
\end{tabular}\\
\begin{tabular}{p{3cm} p{1.75cm} p{1.5cm} p{1.5cm}}
  A=C+v & J$^{\pi}$ & $nlj$ & A  \\
  \hline
\\
  $^{12}$C=$^{9}$Be $\oplus$ $^{3}$He &  0$^{+}$  &  2p$_{3/2}$$^c$ & 1.224$^c$  \\

\\ \hline\hline
\label{parametrospotencialoptico}
\end{tabular}\\
\end{centering}
\begin{tabular}{l}
\hspace{0.0cm} $^{a}$ Ref. \cite{12,13}.\\
\hspace{0.0cm} $^{b}$ Adjusted by FRESCO code to reproduce the cluster  \\
\hspace{0.0cm}  binding energy.\\
\hspace{0.0cm} $^{c}$ Ref. \cite{7,14}.
\end{tabular}\\

\end{table}

The calculations for $^{12}$C($^9$Be,$^{12}$C)$^9$Be elastic
transfer channel were done using the potential obtained in the
optical model analysis. The $^{12}$C was considered as a cluster
structure, composed of a $^9$Be core and a $^3$He valence particle
in a single-particle state. The bound state wave functions were
generated using a binding potential with a Woods-Saxon shape, with
geometric parameters shown in table \ref{tabela2}. The depth was
adjusted to give the correct separation energy of the clusters.

In the calculation, $^3$He is considered to be transferred from the
2p$_{3/2}$ single particle state in $^{12}$C(J$^{\pi}$=0${^+}$) to
the same orbital on the $^9$Be(J$^{\pi}$=3/2${^-}$) projectile. The
spectroscopic factor for the $^{12}$C$_{g.s}$ = $^9$Be$_{g.s}$
$\oplus$ $^3$He cluster structure was taken from the literature and
is listed in table \ref{tabela2}.

The calculated angular distributions are shown in Fig.
\ref{elastico}, as dash-dotted lines, and we can see the fair
agreement with experimental data for 17.3, 19.0 and 21.0 MeV,
showing the importance of the elastic transfer process at these
energies. However, for 13.0 and 14.5 MeV, the importance of the
inclusion of these channels is not clear. In the case of 19.0 MeV,
one can notice that the theoretical prediction presents a good
agreement when compared with the experimental data from Ref.
\cite{7}.

\subsection{Compound Nucleus Formation}

The formation of compound nucleus is another mechanism that can also
contribute to increase the cross sections at intermediate and
forward angles \cite{15,16}. Usually, in this process the nuclei
fuse completely, forming an intermediate state ($^9$Be +$^{12}$C
$\to$ $^{21}$Ne$^{*}$), which after a characteristic time, decays
populating other open channels, including the entrance channel. For
the latter case we have the compound elastic (CE), which for
$^9$Be+$^{12}$C system is reflected in the process $^9$Be +$^{12}$C
$\to$ $^{21}$Ne$^{*}$$\to$ $^9$Be +$^{12}$C.

\begin{table}[h]
\begin{centering}
\caption{Levels considered.} \label{tabela3}
\begin{tabular}{p{3cm} p{4cm} p{3cm}}
  \hline\hline
  Residual Nucleus & Level density parameter (MeV$^{-1}$)$^{a}$ & Number of Discrete levels \\
  \hline
  $^{20}$Ne &  0.16  & 10     \\

  $^{20}$F &  0.16  & 11     \\

  $^{13}$C &  0.16  & 11     \\

  $^{18}$O &  0.16  & 10     \\

  $^{17}$O &  0.16  & 10     \\

  $^{12}$C &  0.16  & 5      \\
  \hline\hline

\end{tabular}\\
\end{centering}

\begin{tabular}{l}
\hspace{0.0cm} $^{a}$ Ref. \cite{16}.\\
\end{tabular}\\

\end{table}

The calculation was performed using the Hauser-Feshbach STATIS code
\cite{17}. The nuclear level density has been described by means of
a level density expression given by Lang \cite{18}, and the
transmission coefficients were determined by an internal Fermi
parametrization \cite{17}. The levels considered in this
calculations are listed on table \ref{tabela3} and are quite similar
to those used in the Ref. \cite{16}.

\begin{figure}[htb!]
\centering
\includegraphics[scale=0.8]{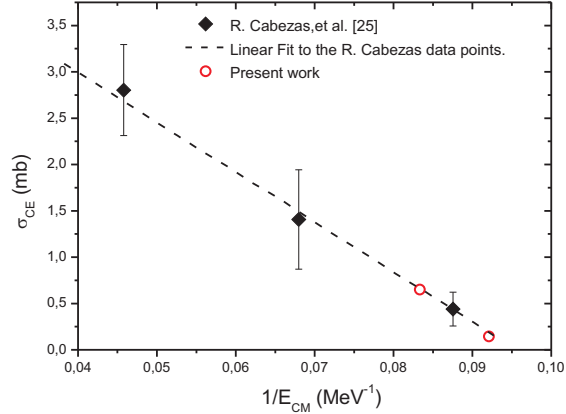}
\caption{ Excitation function for Compound Nucleus Formation. The
open circle points are to E$_{Lab}$ = 19.0 and 21.0 MeV. The
calculated points, were normalized to agree with behavior of data
from \cite{15} (Color Online).} \label{secaototalcompound}
\end{figure}

\begin{figure}[htb!]
\centering
\includegraphics[scale=0.7]{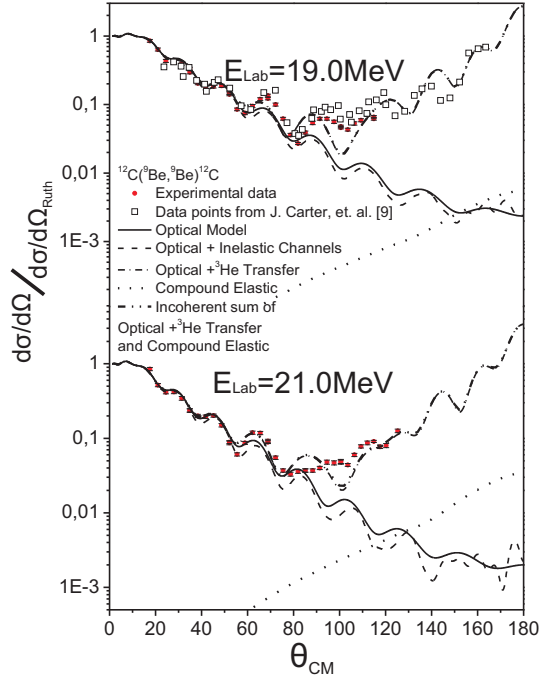}
\caption{Angular distributions for the $^9$Be + $^{12}$C system at
E$_{Lab}$ = 19.0 and 21.0 MeV. The dot lines are the CE angular
distributions from HF calculations. We can notice that the
contribution of CE to differential cross section enhancement at
backward angles is not significante (Color Online).} \label{total}
\end{figure}

Finally the angular distributions are normalized in such a way that
the total cross section agree with the results presented in the
excitation function from Ref. \cite{15} (Figure
\ref{secaototalcompound}). When the obtained angular distributions
are incoherently added to the $^{3}$He transfer results, we can see
that the contribution of this reaction mechanisms is not relevant,
even for our highest energies, as shown in Fig. \ref{total} for the
two highest energies E$_{Lab}$ = 19.0 and 21.0 MeV.

\newpage
\section{Conclusions}

In this work we measured elastic scattering angular distributions
for the $^9$Be+$^{12}$C light system at bombarding energies ranging
from 13.0 MeV to 21.0 MeV. The double folding S\~ao Paulo potential
was used in the analysis that was performed in four steps.

In the first one we considered N$_{r}$ and N$_{i}$ as free
parameters for the fits to the angular distributions at the forward
angular region. The angular distributions calculated with the
optical model have shown a good agreement with the 13.0 and 14.5 MeV
experimental data. For 17.3, 19.0 and 21.0 MeV the agreement is
reasonable at forward angles. However, the description at backward
angles is not good, suggesting that the coupling to other mechanisms
is important.

In the second step of the analysis, no evidence of the coupling to
inelastic channels was observed. Using the spectroscopic factors
extracted from the literature for $^{12}$C=${^3}$He$\oplus$
${^9}$Be, in the third step we took into account the ${^3}$He
elastic transfer channel. For the ${^3}$He transfer we see a
pronounced improvement in the description of the data for 17.3, 19.0
and 21.0 MeV, which suggests that this process is important at
intermediate and backward angles. Finally in the fourth step,
corresponding to the compound elastic calculation, the results
suggest that this mechanism is not an important process at the
energies studied in this work.

The energy dependence of the N$_r$ parameter suggest the presence of
the threshold anomaly. However, no strong conclusions could be made
due to the constant value of N$_i$.

Finally, to obtain information about the effects of the elastic
transfer process and check the values of spectroscopic factors, it
would be important to perform a more carefully analysis and
measurements of the elastic scattering angular distributions at more
backward angles.












\end{document}